\documentclass[aps,a4paper,10pt,twocolumn,nofootinbib,eqsecnum]{revtex4} 
\usepackage[T1]{fontenc}
\usepackage{mathptmx}
\usepackage{datetime}
\usepackage{amsmath}
\usepackage{amsfonts}
\usepackage{mathrsfs}
\usepackage[mathscr]{euscript}
\usepackage{mathtools}
\usepackage{textcomp} 
\usepackage{fancyhdr}
\usepackage{colordvi}
\usepackage{hyperref} 
\usepackage{epsfig}
\usepackage{color}
\usepackage{bm}

\def\overstrike#1#2{{\setbox0\hbox{$#2$}\hbox to \wd0{\hss
    $#1$\hss}\kern-\wd0\box0}}

        \DeclareMathOperator{\grad}{\nabla}

\renewcommand{\Vec}[1]{\textbf{#1}}

\newdateformat{yymmdddate}{\THEYEAR/\twodigit{\THEMONTH}/\twodigit{\THEDAY}}

\begin{document}

\title{Causal diagrams for physical models}


\author{Paul Kinsler}
\email{Dr.Paul.Kinsler@physics.org}


\affiliation{
  Blackett Laboratory, Imperial College London,
  Prince Consort Road,
  London SW7 2AZ,
  United Kingdom.}

\begin{abstract}

I present a scheme of drawing causal diagrams
 based on physically motivated mathematical models
 expressed in terms of temporal differential equations.
They provide 
 a means of better understanding the processes and causal relationships
 contained within such systems.

\end{abstract}

\lhead{\includegraphics[height=5mm,angle=0]{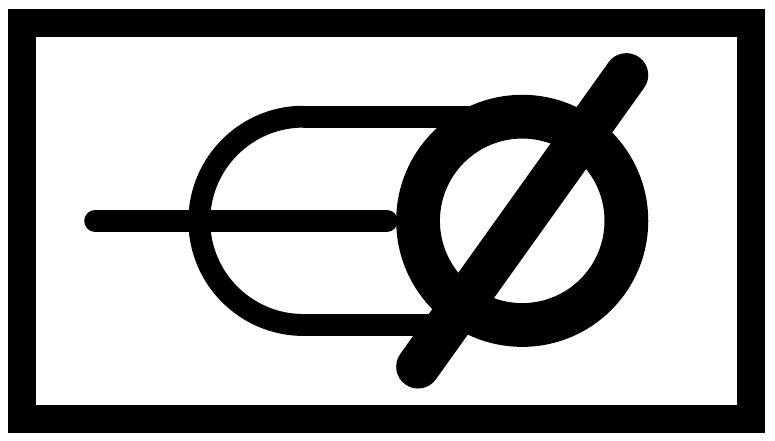}~~DICAUS}
\chead{Causal Diagrams}
\rhead{
\href{mailto:Dr.Paul.Kinsler@physics.org}{Dr.Paul.Kinsler@physics.org}\\
\href{http://www.kinsler.org/physics/}{http://www.kinsler.org/physics/}
}

\date{\today}
\maketitle
\thispagestyle{fancy}


\section{Introduction}
\label{S-Introduction}

This paper presents a systematic way 
 of making diagrammatic representations 
 of causal physical models.
The premise follows that due to 
 the presentation of Kinsler \cite{Kinsler-2011ejp}, 
 with a notion of causality being linked
 to the existence of temporal derivatives 
 in physical models based on differential equations.
Whilst many other approaches to causality are both useful
 and relevant to many areas, 
 the one here is chosen to be simple and direct:
 it analyses the causal properties of an existing model, 
 and acts as a guide for the  construction
 of models consistent with its principles.

The original motivation for these diagrams 
 was in analogy with the fishbone-like Ishikawa diagrams
 \cite{Wiki-Ishikawa}, 
 used in business, 
 and created by Kaoru Ishikawa 
 in order to identify the origins of a particular effect.
However, 
 unlike in business where the analysis, 
 no matter how carefully undertaken, 
 is far from being completely specified, 
 in physical diagrams I can add features
 to precisely represent the physical model.
Note that a simple way to ``physicsify'' Ishikawa diagrams
 had been implemented by Thomas Wong \cite{Essay3YP-Causality-Wong}, 
 in a third year undergraduate essay project at Imperial College London
 which I proposed and then supervised in 2015.

The diagramming process introduced here is distinct from
 that used in 
 standard block diagrams \cite{WikiBooks-BlockDiagrams}.
In block diagrams, 
 elements have specified and \emph{defined} inputs and outputs; 
 whereas in my scheme each node indicates what \emph{modifies}
 the quantity of interest.
Further, 
 block diagrams emphasizes solutions (e.g. by representing integrators)
 or frequency domain properties; 
 whereas my aim is to emphasize on-going dynamical behaviour.
In particular, 
 block diagrams often loop outputs back to inputs to represent feedback, 
 whereas here such loop-like constructions would 
 represent non-causal behavior.

Another type of diagrams used in causal analyses are
 ``directed acyclic graphs'' (DAGs), 
 and in addition to more technical aspects \cite{MathWorld-AcyclicDigraphs}
 have been proposed as a methodology to assist
 draw causal inferences from non-experimental data
 (see e.g. \cite{Elwert-GCM}).
When applied in the contexts for which they are intended, 
 which is not usually for physical modelling, 
 they are a valuable tool
 for investigating and clarifying causal relationships.
One of the promotional points about DAG's 
 is that they are a mathematics-free 
 method that relies on graphical rules for their construction.
However, 
 as a theoretical physicist, 
 I would prefer any graphical rules
 to \emph{also} be a re-representation 
 of a specific mathematical model -- 
 and this indeed is the case for the diagrams I propose here.
Thus my proposed diagrams do not only say ``this causes an effect on that'', 
 but also contain a mathematical specification of it.
Nevertheless, 
 one might start by constructing a diagram for a causal physical model
 in a DAG-like way,
 based on expectations of cause and effect.
Then, 
 subsequently, 
 this could be augmented with the diagrammatic notation 
 introduced here to specify it more exactly; 
 although note that DAGs whilst can be (and are) 
 explicitly associated with a labelling of cause and effect,
 they do not have the same sense of temporal \emph{dynamics}
 as do many physical theories.
For example, 
 a causal DAG has no way, 
 without an augmentation such as the addition of temporal labelling,
 of diagramming an oscillatory system.


In section \ref{S-core} I present the basic core diagrammatic language, 
 consisting of arrows and dots
 and how they relate to simple temporal differential equations.
This is followed in section \ref{S-combination} 
 by extensions allowing multipliers, 
 parameters, 
 and other operators to be added to the diagrams
 by adding labelled boxes, 
 as well as how quantities that affect themselves should be shown.
In section \ref{S-spatials}
 I add an optional arrowhead notation to simplify the inclusion 
 of spatial derivatives,
 and remark on how simple second order wave equations 
 can be drawn.
After that, 
 in section \ref{S-exmplewaves}
 I show and discuss interconnected systems such as wave models,
 and in particular how the macroscopic Maxwell's equations 
 can be diagrammed.
Finally, 
 I conclude in section \ref{S-conclusions}.


\section{Core diagrams}
\label{S-core}

As defined by Kinsler \cite{Kinsler-2011ejp}, 
 the simplest possible causal model for the behaviour of 
 some quantity $R$ under the influence of a stimulus $Q(t)$,
 is the temporal differential equation
~
\begin{align}
  \frac{dR}
       {dt}
=
  \partial_t R
&=
  Q(t)
,
\end{align}
 although we might also straightforwardly generalize this to
 models with $n$-th order time derivatives as
~
\begin{align}
  \frac{d^n R}
       {dt^n}
=
  \partial_t^n R
&=
  Q(t)
.
\end{align}
Note that even though here $Q(t)$ is some function of all time, 
 and might even have been defined in advance in some 
 pre-determined way, 
 the current value of $R$ is \emph{only} determined by
 the current or past values of $Q$.
In more complicated models, 
 it may also be that the stimulus $Q$ is (or is calculated from)
 some other quantity which has its own dynamical behaviour.

\begin{figure}[ht]
  \begin{center}
    \resizebox{0.45\columnwidth}{!}{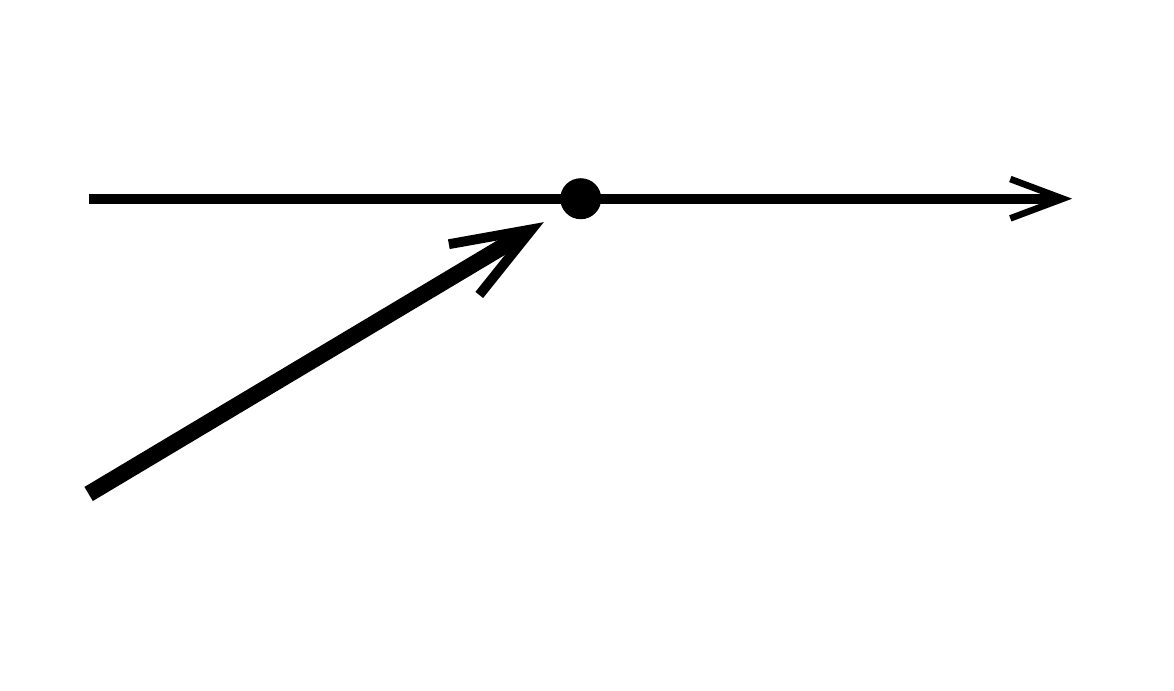}
    \resizebox{0.45\columnwidth}{!}{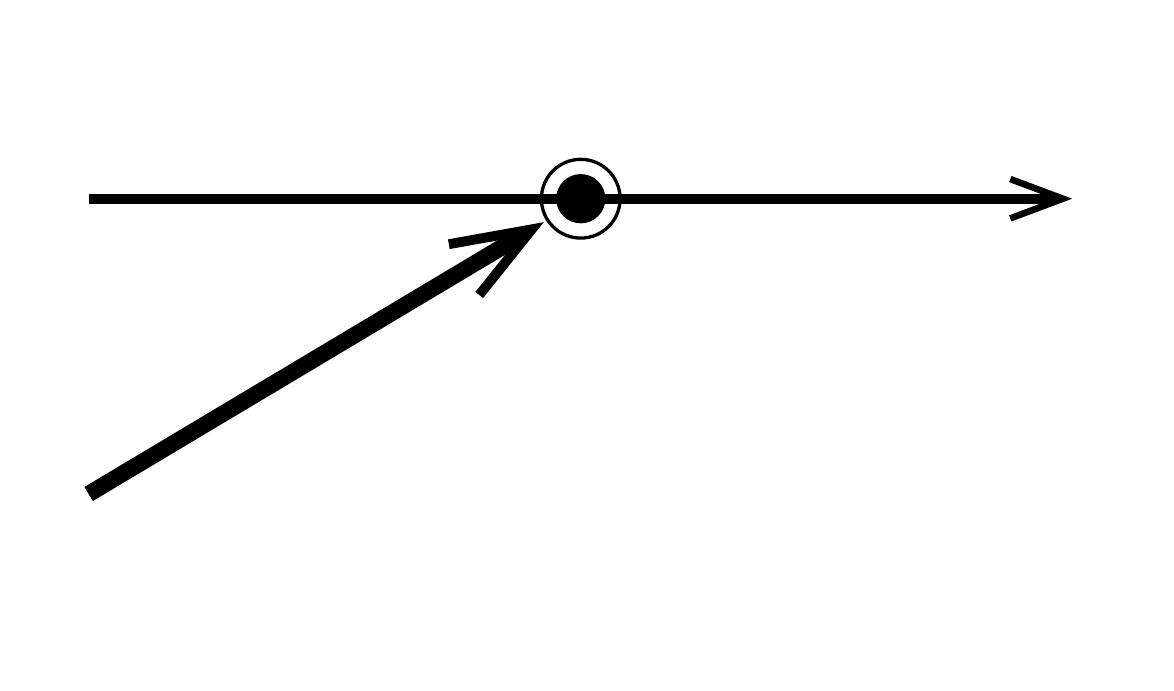}
  \end{center}
\caption[]
{ \label{fig-simple}
Primitive vertices and their notation:
 simple equations with only a first
 or a second order time derivative.
}
\end{figure}

For these simple cases, 
 the core diagrams denote each causal relationship
 by means of a simple vertex with a dot or filled circle.
We denote any quantities that are evolving in time, 
 i.e. that are being affected
 with horizontal lines
 going from left to right.
Other stimuli (causes) that affect it 
 come in as diagonal arrows pointing 
 either up-right or down-right.

In fig. \ref{fig-simple}
 we show the most primitive vertex
 with a known stimulus $Q$
 driving first-order temporal changes in $R$.
If the model has higher order time derivatives, 
 the ``dot'' vertex is augmented by one extra circle
 per extra time derivative.

It is important to note that the time derivative or derivatives
 are only denoted by the dot
 and its surrounding circles (if any),
 and \emph{not} {by the arrowhead}.
In fact, 
 from a purely diagrammatic perspective we could dispense
 with the arrowhead entirely.
However it helps reinforce the intent that time is advancing 
 as we move from left to right across the page, 
 and that some stimuli (written $Q$ in the above)
 is applied and drives changes in a quantity of interest ($R$ above).

%
\section{Scalings and combinations}
\label{S-combination}

\begin{figure}[ht]
  \begin{center}
    \resizebox{0.325\columnwidth}{!}{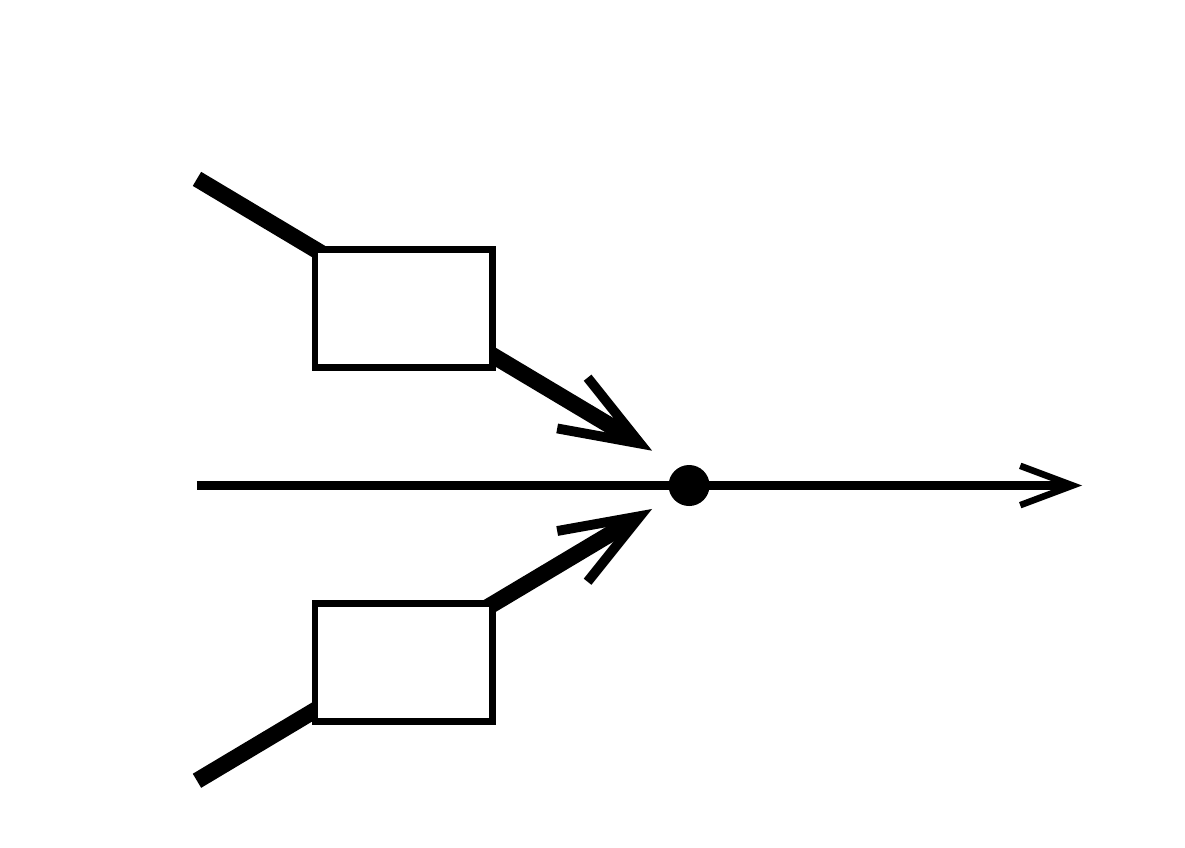}
    \resizebox{0.325\columnwidth}{!}{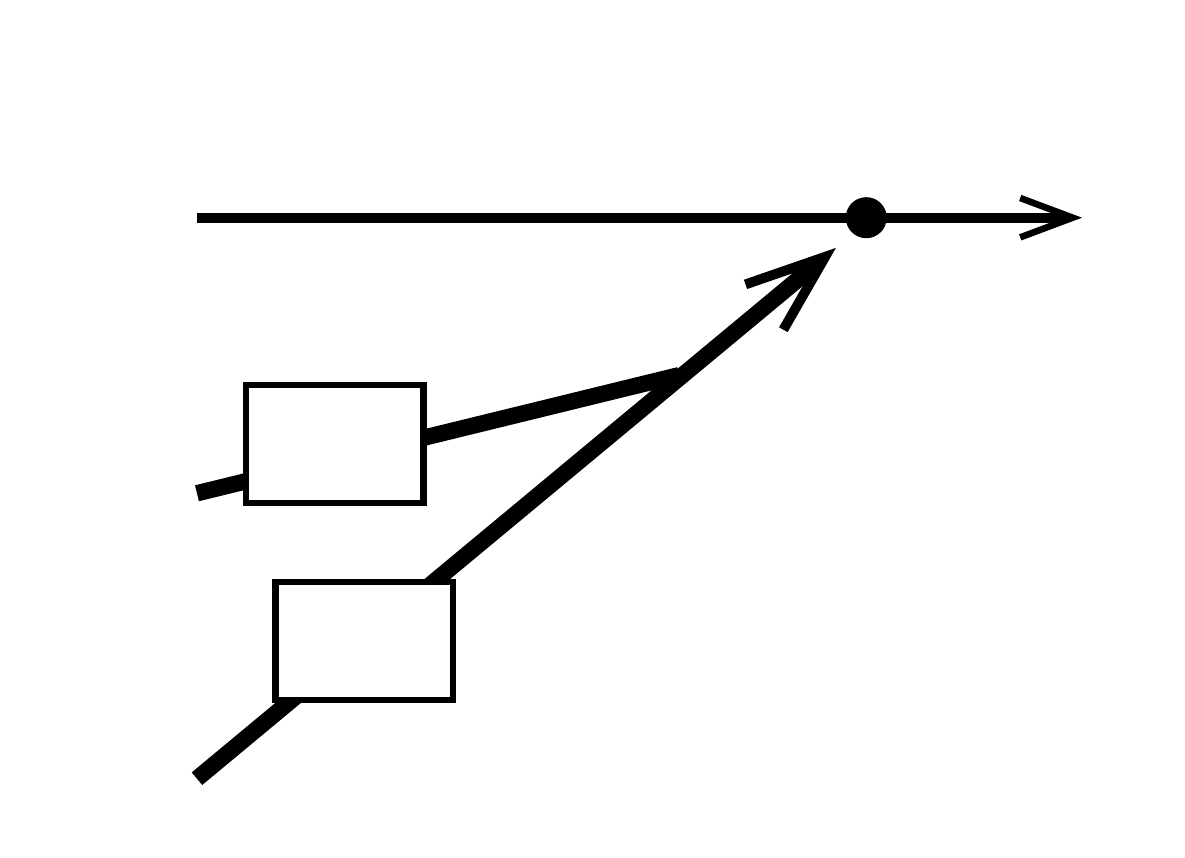}
    \resizebox{0.325\columnwidth}{!}{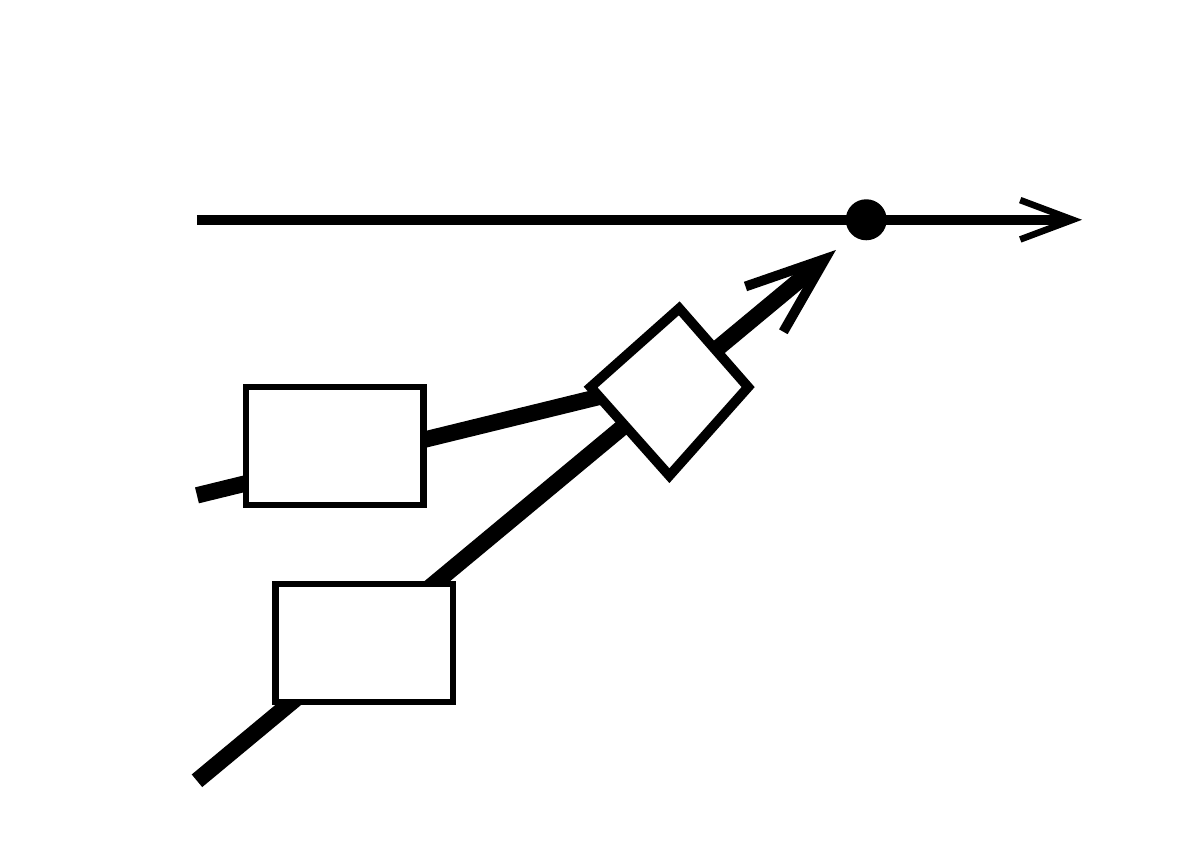}
  \end{center}
\caption[]
{ \label{fig-scaleetc}
Scaling and combination of causes.
The left hand diagram shows the two causes $S$ and $Q$ 
 with their rescaling or functional dependence
 indicated in their respective boxes each on different arrows.
The centre diagram shows the same situation as the left, 
 but with the modified causes $-\mu S$ and $f(Q)$
 added together
 (shown by the \emph{joining} of the the two lines)
 before application to $R$ as indicated by the ordinary arrowhead.
The right hand diagram again re-represents the left hand one, 
 but also introduces the \emph{diamond} element 
 as a way of introducing a combined stimulus $T$.
Although not strictly necessary here, 
 it is an option which can be useful in more complicated diagrams.
}
\end{figure}

Sometimes --
 or indeed typically --
 the various ``cause'' stimuli we might use
 are scaled or changed versions of other quantities or stimuli
 in the model, 
 so on fig. \ref{fig-scaleetc}
 we show how to diagram such cases.
Note that the box notation used here 
 could also be used instead of the varied arrowheads 
 suggested below as a shorthand for common spatial derivative terms.

For a a system influenced by two different stimuli, 
 e.g. $S$ and $Q$, 
 the causal diagram is shown on fig. \ref{fig-scaleetc}, 
 where
 we just put the mathematical terms in a box
 where they can be easily read and understood.
We might say that these boxed quantities
 tell us how the stimuli/causes $S, Q$ are ``conditioned''
 before altering the behavior of the quantity $R$.
While these conditioning (boxed) terms are, 
 in terms of our model, 
 completely specified, 
 the physical processes they represent are very likely
 to be very complicated.
If we were to develop a causal model that explained the $f(Q)$ term 
 in fig. \ref{fig-scaleetc}, 
 for example, 
 it may well generate a complicated interconnected mesh-like diagram
 involving a variety of relevant quantities and/or stimuli.

Note that to avoid ambiguity
 it is necessary to put the full expression 
 in each conditioning box.
For example, 
 in fig. \ref{fig-scaleetc}, 
 we must write $-\mu S$ in the box(es) conditioning $S$.
If we were to put only $-\mu$ and \emph{assumed} that
 a multiplication by $S$ was implied, 
 an alternate conditioning term $-\mu {U}^2$ derived from a stimulus $U$
 would require us to put the potentially confusing $-\mu U$ in the box.

\begin{figure}[ht]
  \begin{center}
    \resizebox{0.45\columnwidth}{!}{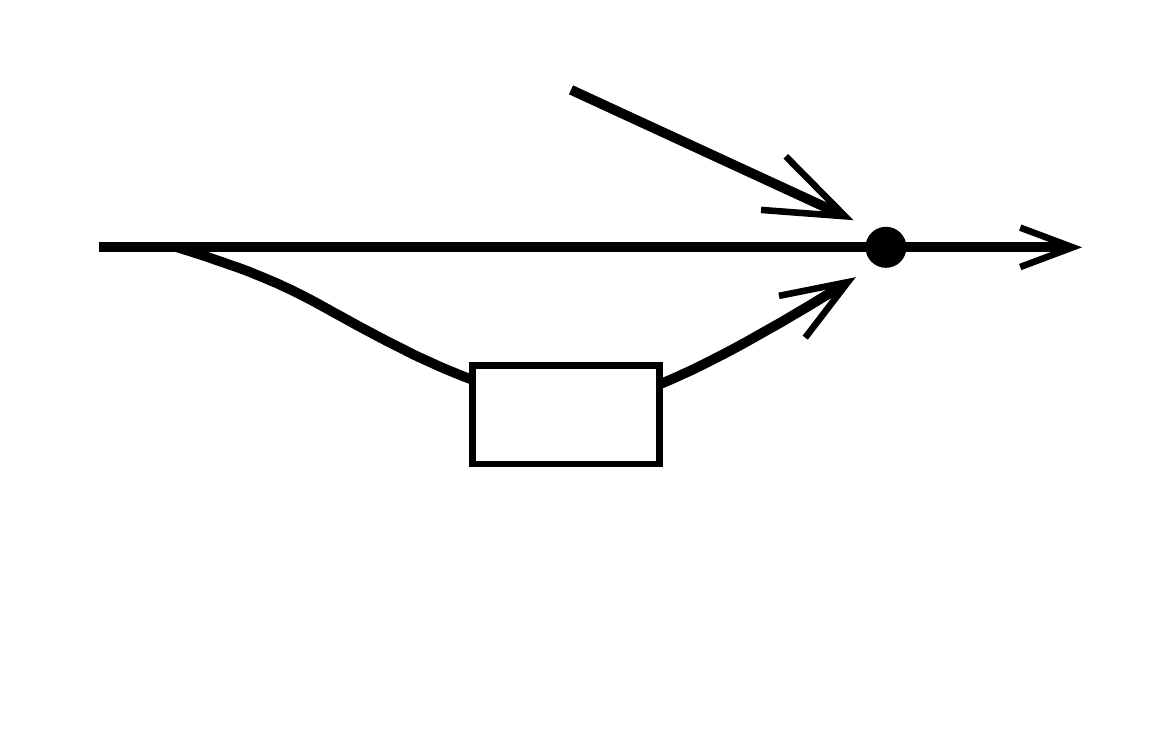}
    \resizebox{0.45\columnwidth}{!}{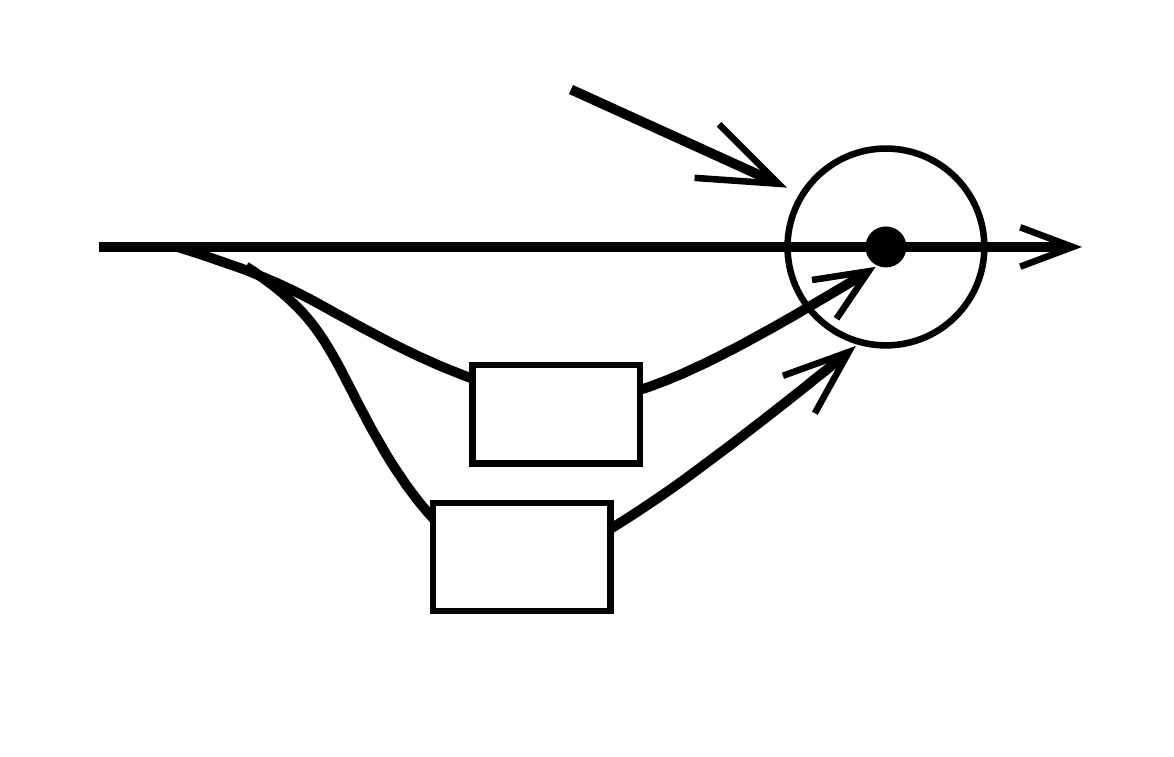}
  \end{center}
\caption[]
{ \label{fig-selfdecay}
\textbf{Left:} 
 Simple linear decay, 
 with a driving term $F$
 and a self-acting term representing a loss
 proportional to $\gamma$. ~~
\textbf{Right:}
Simple Lorentz oscillator model,
 with a driving term $F$
 and a self-acting term representing a loss
 proportional to $\gamma$, 
 and a natural resonance frequency $\omega_0^2$.
Where each arrow terminates is important:
 the loss term is a first derivative effect and so its arrow
 ends on the central dot; 
 the driving and resonant frequency terms are second derivative
 and terminate on the circle.
}
\end{figure}

Many quantities of interest 
 have behaviour that depends on their current state --
 a simple driven, 
 linearly damped system is one such.
For a loss parameter $\gamma$ and driving term $F$, 
 the causal diagram is shown on 
 the left hand side of fig. \ref{fig-selfdecay}.
The Lorentz oscillator model in fig. \ref{fig-selfdecay}   
 is a little more complicated.
It has two parts -- 
 a first derivative ``loss'' part 
 with $\partial_t R \propto -\gamma R$
 needs to be added to a second derivative ``oscillation \& driving'' part
 with 
  $\partial_t^2 R \propto - \omega_0^2 R + F$.
To do this we turn the first part also into a second derivative; 
 i.e. $\rightarrow \partial_t^2 R \propto -\gamma \partial_t R$, 
 to give the expected (Lorentz) sum, 
~
\begin{align}
  \partial_t^2 R 
&=
 -
  \gamma \partial_t R
 -
  \omega_0^2 R
 +
  F
.
\end{align}
The outer circle, 
 denoting the second derivative effects, 
 is here drawn much larger than in fig. \ref{fig-simple}
 because of the need to visually distinguish between 
 the effect of the loss term whose net effect is first order 
 ($\sim\partial_t R$)
 and the oscillation and driving terms whose net effects
 are second order
 ($\sim\partial_t^2 R$).

\begin{figure}[ht]
  \begin{center}
    \resizebox{0.45\columnwidth}{!}{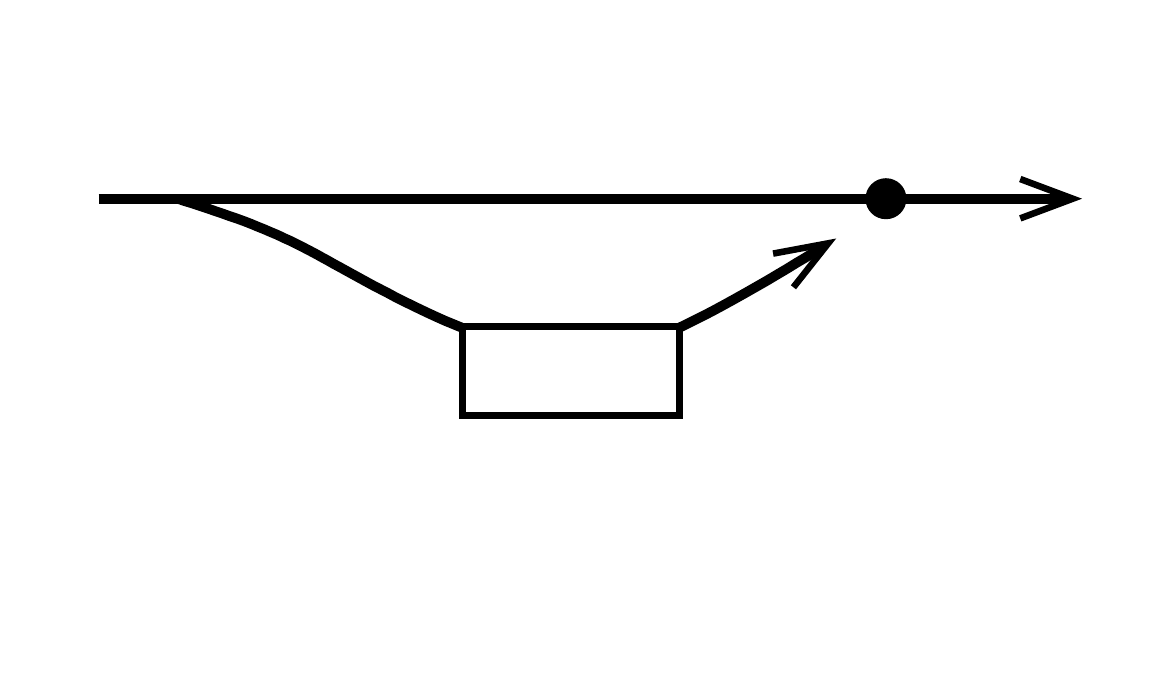}
    \resizebox{0.45\columnwidth}{!}{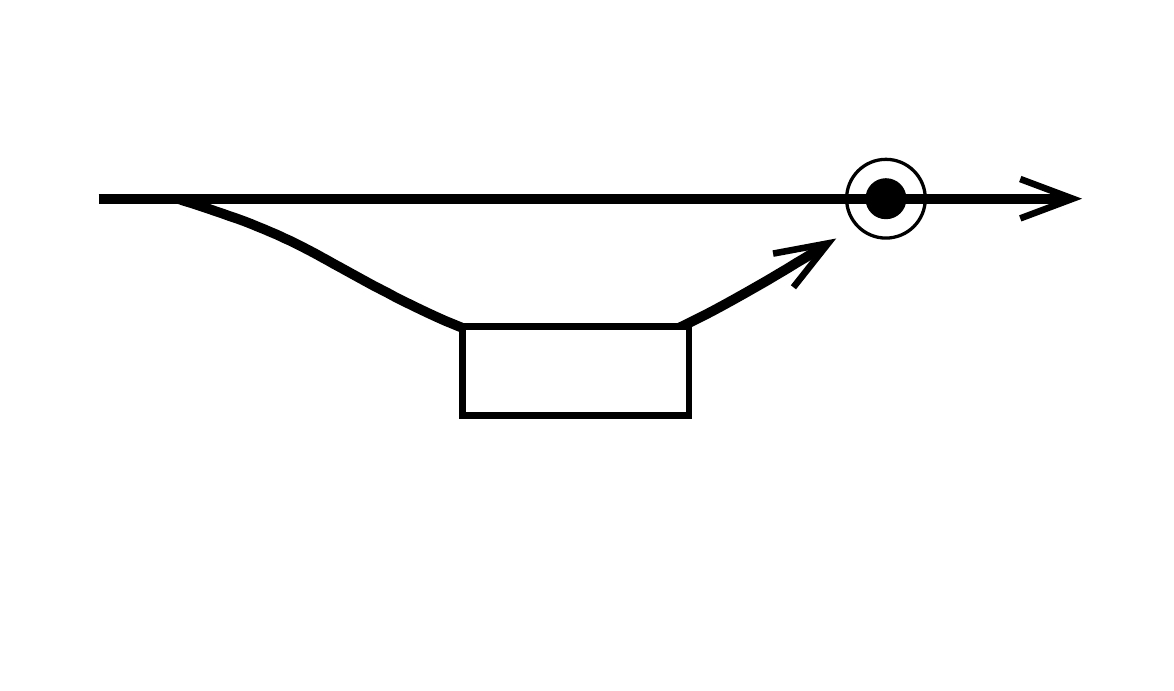}
  \end{center}
\caption[]
{ \label{fig-selfwave}
Simple single-field second order wave equations,
 where a scalar field quantity $R$ acts on itself after conditioning: 
 \emph{variation} in its spatial profile 
 causes changes in the temporal evolution.
On the left hand side we have diagrammed 
 a diffusion or Schr\"odinger-like equation, 
 with their characteristic single time derivatives, 
 whereas on the right hand side 
 we have a Helmholtz-like equation.
As usual, 
 the spatial gradient operator is defined with
 $\grad = (\partial_x, \partial_y, \partial_z)$.
}
\end{figure}

Other systems we might make diagrams for are simple waves.
In a scalar wave, 
 the temporal response of the wave amplitude 
 depends on the spatial changes in profile of the wave.
Two common cases are ones with first order time derivatives, 
 such as the Schr\"odinger wave equation, 
 and those with second order time derivatives, 
 such as the Helmholtz equation.
Shorn of any parameters such as wave speeds
 in order to simplify the diagrams, 
 these are shown on fig. \ref{fig-selfwave}.

%
\section{Spatial derivatives}
\label{S-spatials}

\begin{figure}[ht]
  \begin{center}
    \resizebox{0.32\columnwidth}{!}{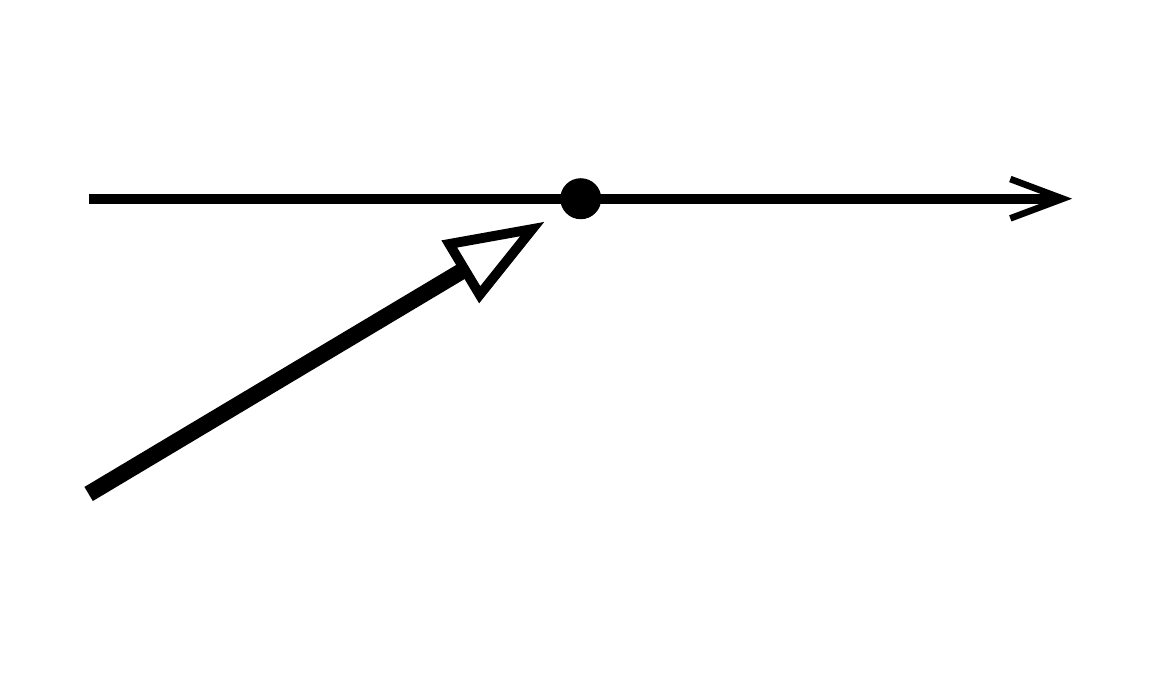}
    \resizebox{0.32\columnwidth}{!}{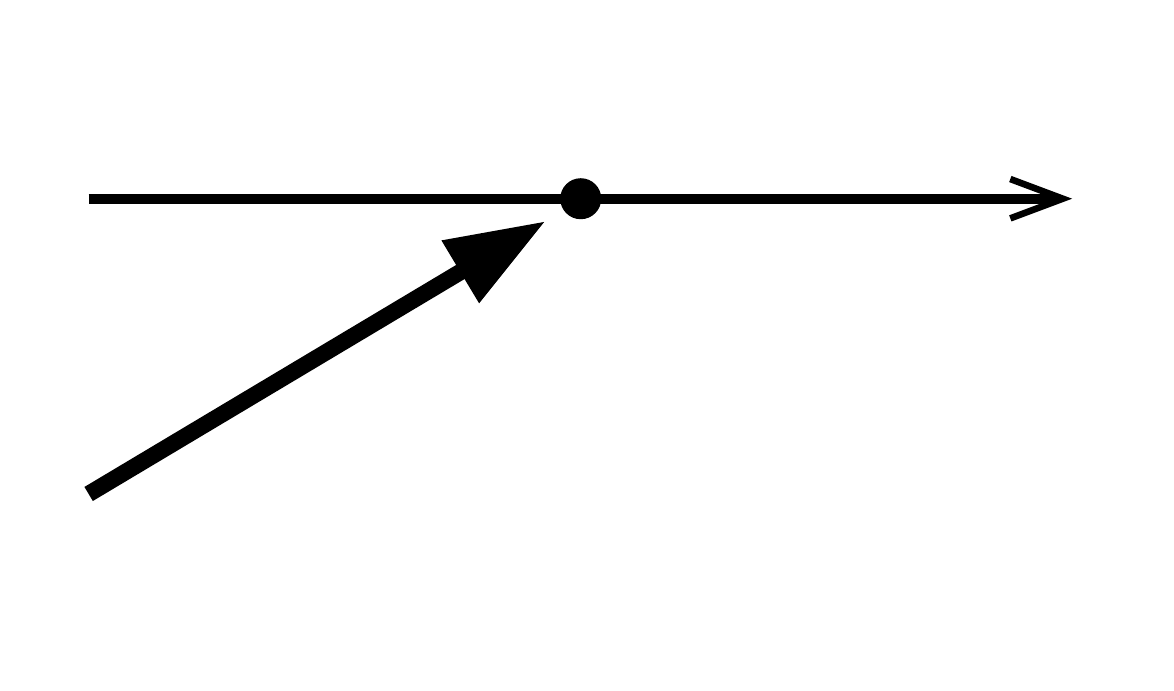}
    \resizebox{0.32\columnwidth}{!}{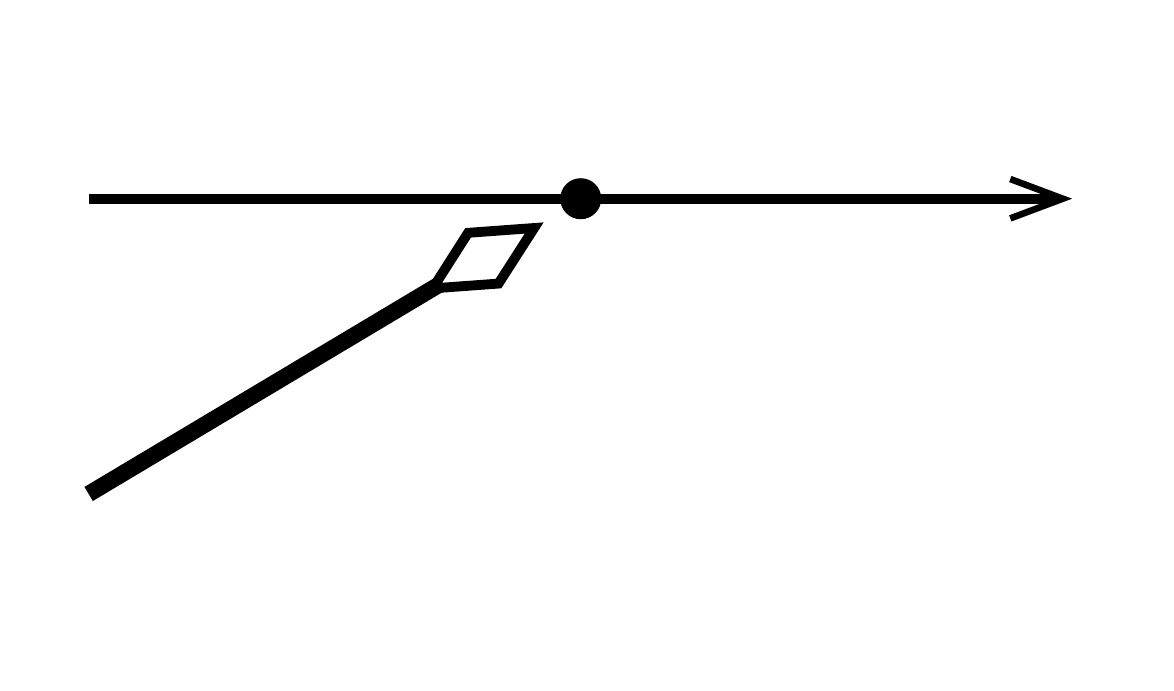}
  \end{center}
\caption[]
{ \label{fig-dx}
Simple vertices involving spatial derivatives; 
 with an open arrowhead for the usual gradient ($\grad$), 
 a closed arrowhead for the divergence ($\grad \cdot$), 
 and a diamond arrowhead for the curl ($\grad \times$).
Here, 
 a boldface symbol denotes a vector quantity.
}
\end{figure}

As we saw in the latter part of the previous section, 
 and indeed in many interesting physical systems, 
 the ``cause'' affecting the quantity of interest
 is a spatial derivative of some kind.
It is therefore useful to utilize a shorthand notation
 of different arrowhead types to denote 
 the three vector calculus spatial derivatives of interest, 
 which we denote by using different kinds of arrowhead, 
 as shown on fig. \ref{fig-dx}.

\begin{figure}[ht]
  \begin{center}
    \resizebox{0.32\columnwidth}{!}{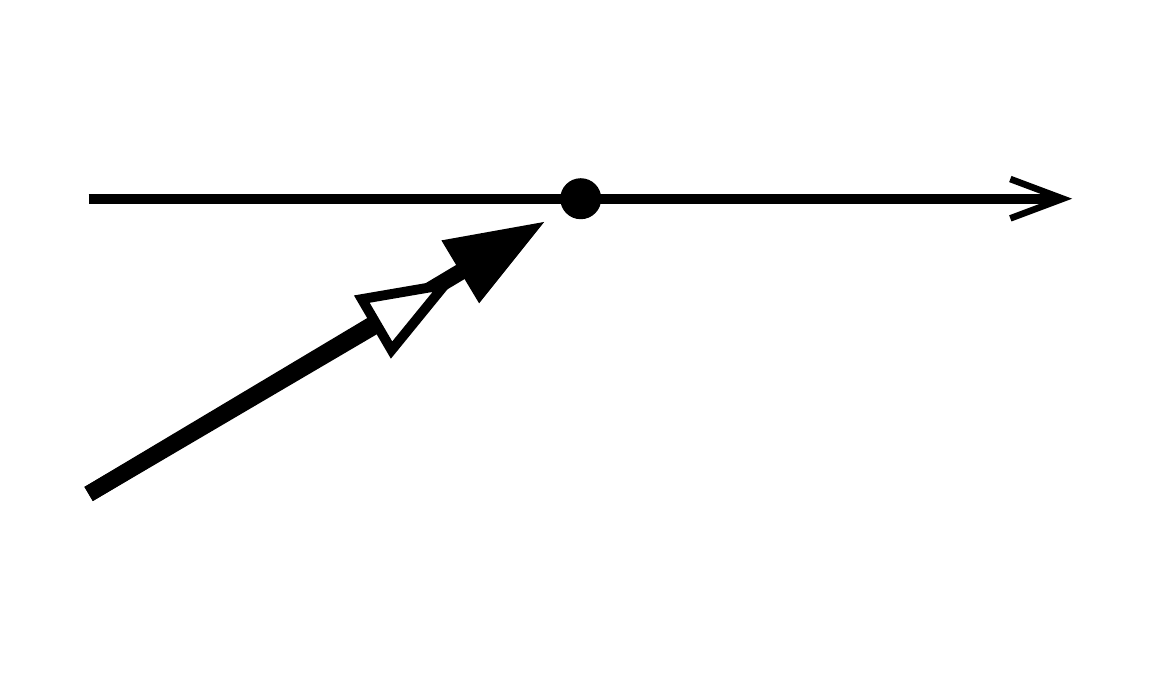}
    \resizebox{0.32\columnwidth}{!}{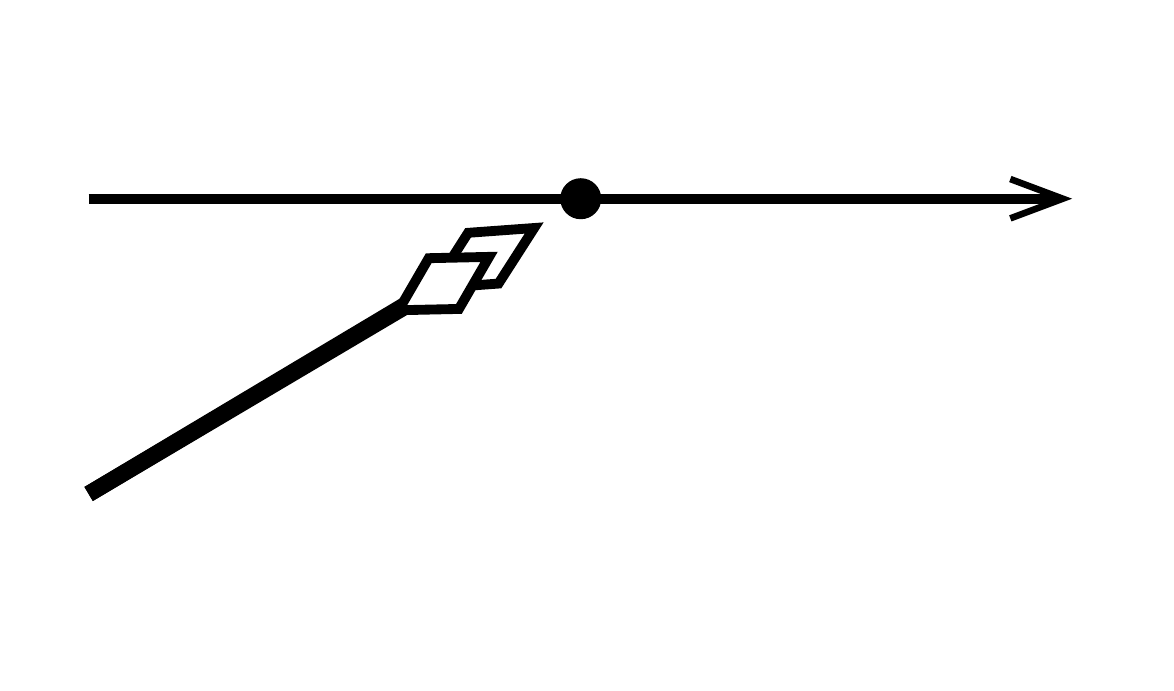}
    \resizebox{0.32\columnwidth}{!}{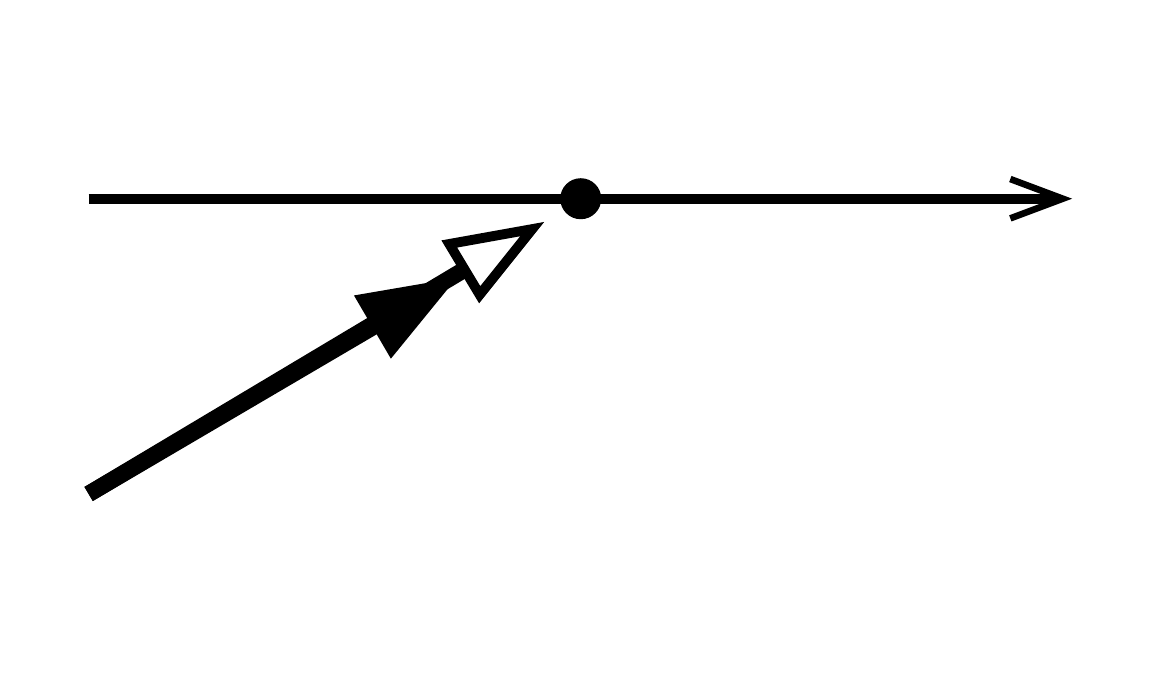}
  \end{center}
\caption[]
{ \label{fig-d2x}
Vertices involving double spatial derivatives.
Note the ordering of the arrowheads, 
 which may at first seem counter-intuitive.
The arrowheads appear along the line towards the dot
 in the order with which they are applied to $Q$, 
 thus in the left hand diagram the open arrowhead (for $\grad$)
 appears first, 
 then the closed arrowhead (for $\grad \cdot$) second.
The reason for this can be seen in the right hand diagram 
 on fig. \ref{fig-wave}.
}
\end{figure}

\begin{figure}[ht]
  \begin{center}
    \resizebox{0.32\columnwidth}{!}{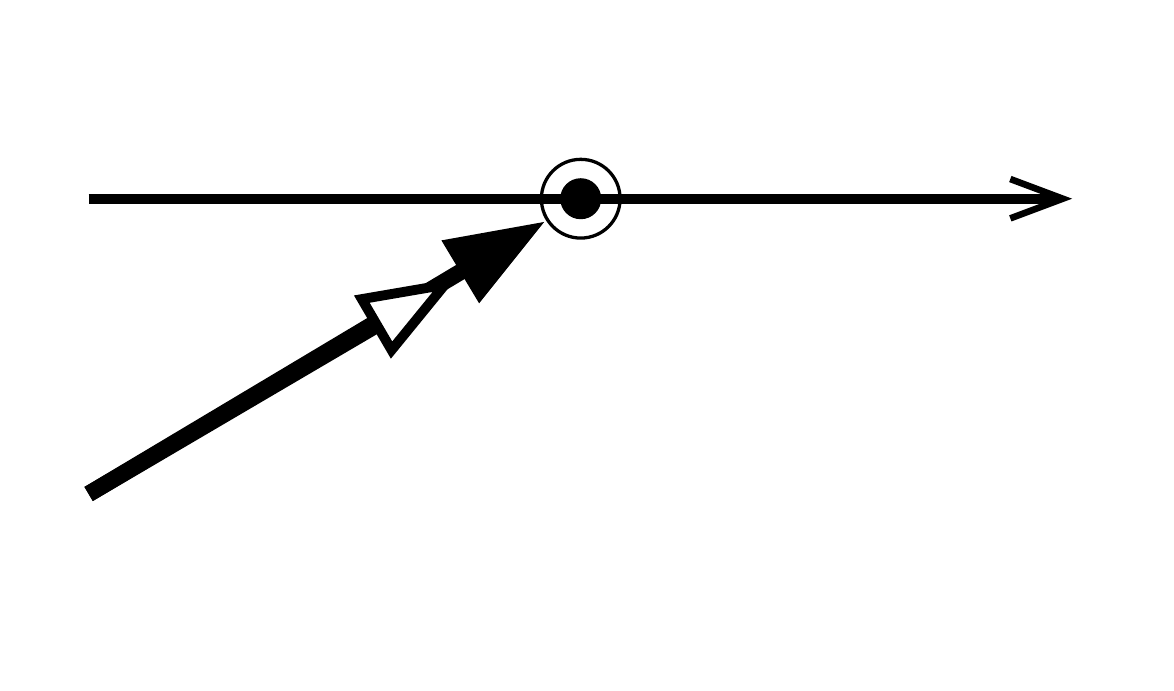}
    \resizebox{0.32\columnwidth}{!}{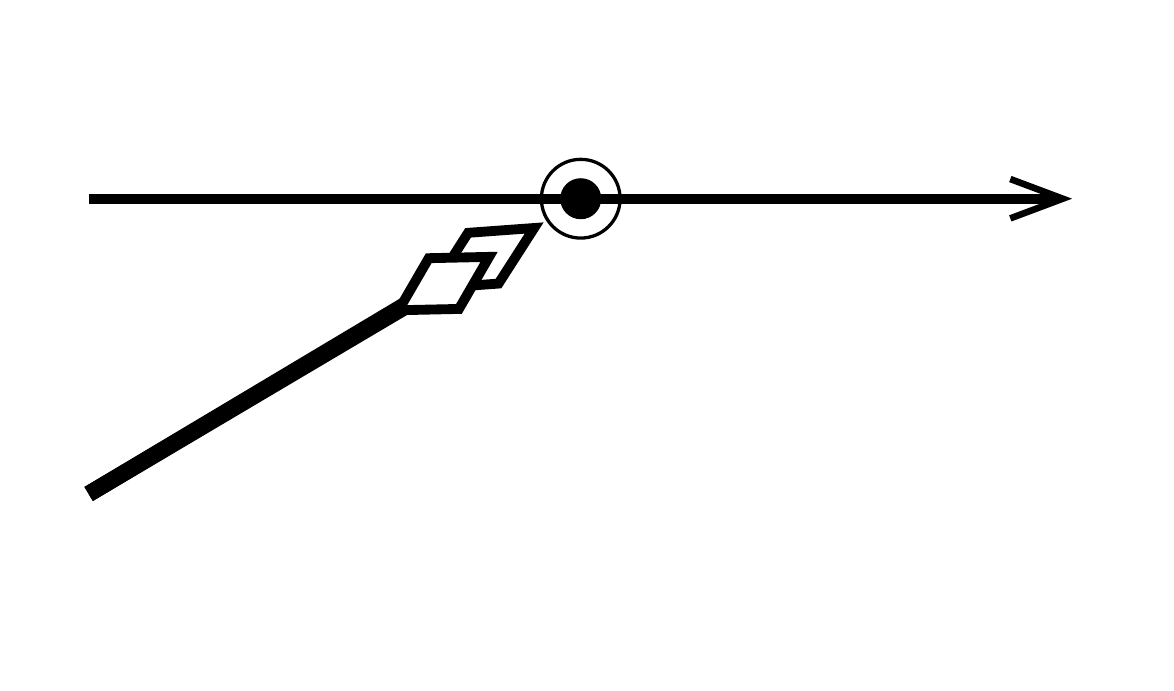}
    \resizebox{0.32\columnwidth}{!}{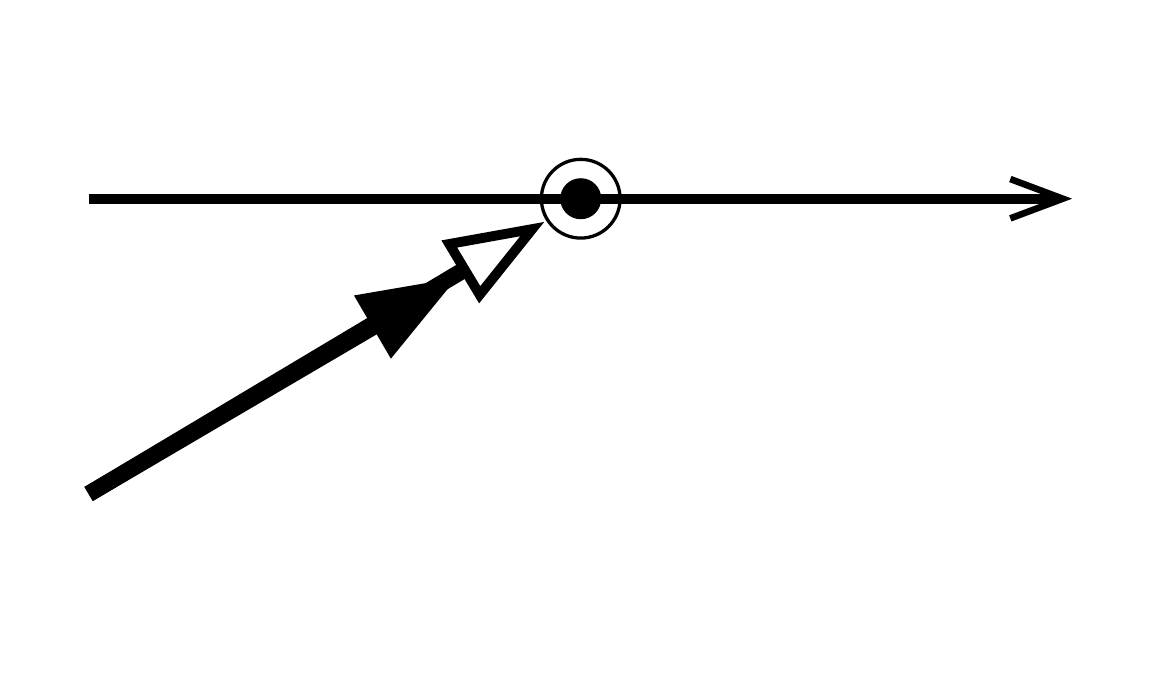}
  \end{center}
\caption[]
{ \label{fig-d2wave}
Vertices involving double spatial derivatives 
 driving doubled time derivatives.
}
\end{figure}

\begin{figure}[ht]
  \begin{center}
    \resizebox{0.325\columnwidth}{!}{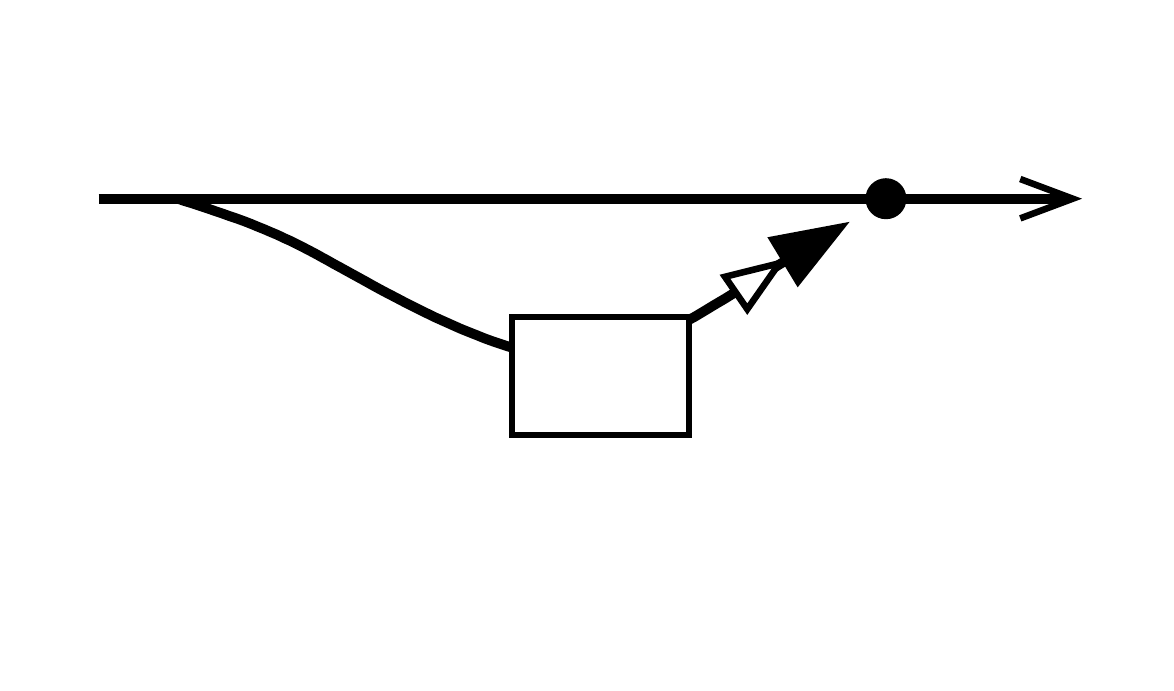}
    \resizebox{0.325\columnwidth}{!}{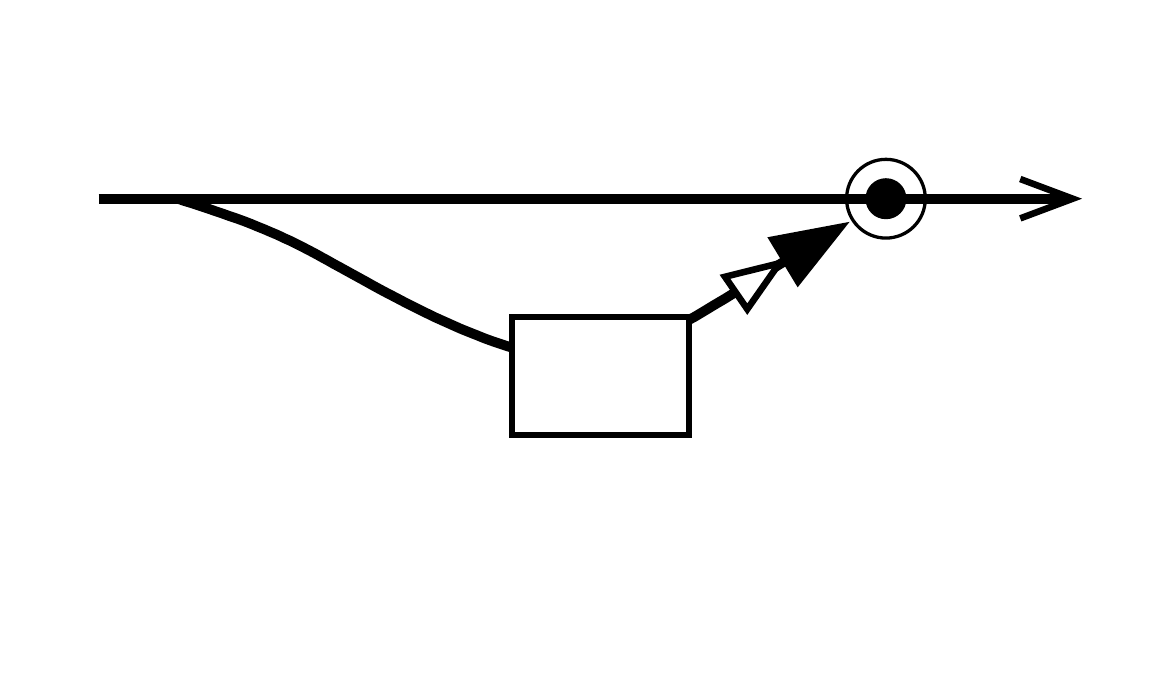}
    \resizebox{0.325\columnwidth}{!}{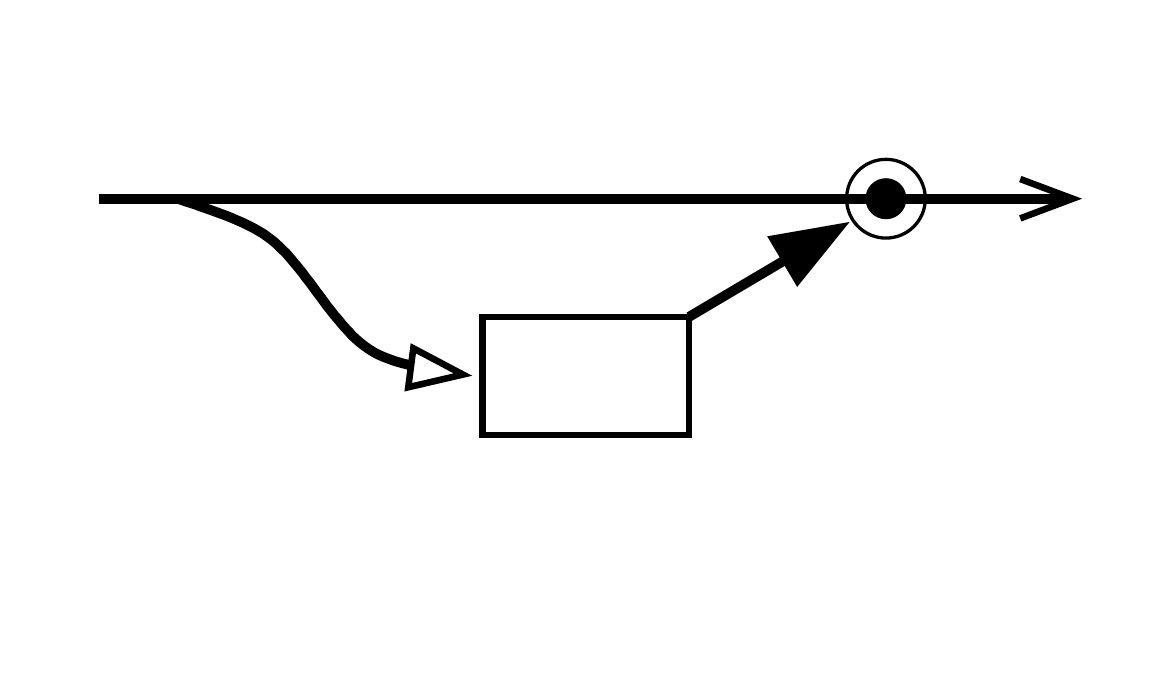}
  \end{center}
\caption[]
{ \label{fig-wave}
The wave diagrams for fig. \ref{fig-selfwave}, 
 but using the new arrowhead notations.
Also added are some necessary wave parameters, 
 a length scale $\Lambda$ 
 and a wave speed $c$.
In the right hand diagram we can see the result of moving the $c^2$ parameter
 in between the spatial derivative terms, 
 and by comparing with the centre diagram, 
 also see how the arrowhead ordering has been preserved.
Note that 
 the open arrowhead denoting the gradient $\grad$ 
 might now be considered redundant, 
 since it must also be included in the boxed term; 
 however it does aid the readability of the diagram.
}
\end{figure}

For multiple derivatives, 
 we can simple stack the arrowheads up in the correct order, 
 which is the order in which the operator they represent is applied to $Q$,
 as seen on fig. \ref{fig-d2x} and fig. \ref{fig-d2wave}.
Further, 
 we can re-diagram the wave models of fig. \ref{fig-selfwave}.
The result of this process
 is shown on fig. \ref{fig-wave}, 
 where simple parameters, 
 a wave speed $c$ and a length scale $\Lambda$ are also included.
\emph{Note the ordering of the box and arrowheads
 in relation to the equation.}
As we follow the ``cause'' line, 
 we meet the boxes first
  (so that e.g. $R \rightarrow c^2 R$), 
 and then the derivative arrowheads 
  (so that $c^2 R \rightarrow \grad \cdot \grad (c^2 R)$).

At this point we have all the tools necessary to 
 draw causal diagrams of any physical model,
 at least in principle.
However, 
 in the next section 
 I will apply the concept to some more specific models
 in order to see how the method works out in practice.
One point to note is that is can be tricky, 
 especially in models with multiple quantities being affected
 (``effected''),
 to keep the diagrams well organized enough to remain readable.
While a valuable process in itself, 
 because it forces one to be very clear about cause and effect
 as implemented in the mathematical model, 
 it may be that restricting the use of these diagrams
 to fragments of a larger model is their most practical use.
The challenge results largely from the fact that each quantity
 needs to be represented by a \emph{line} that represents
 its a temporal history.
In contrast, 
 non-dynamical diagramming schemes like DAGs or block diagrams, 
 which represent relevant quantities as discrete localized elements
 such as letters or blocks, 
 allow greater freedom when drawing multiple interconnections.

%
\section{Interconnected processes}
\label{S-exmplewaves}


Many physical systems have more than a single quantity 
 undergoing changes due to their environment; 
 thus, 
 in our diagrams, 
 we may need more than one horizontal line.
One notable type of interconnected system is a \emph{wave}, 
 where (e.g.)
 the displacement and speed profiles of a stretched string
 are coupled together
 in a way that forms oscillations that travel along it.
Waves are therefore an ideal test case 
 for the diagramming of interconnected systems.

\begin{figure}[ht]
  \begin{center}
    \resizebox{0.45\columnwidth}{!}{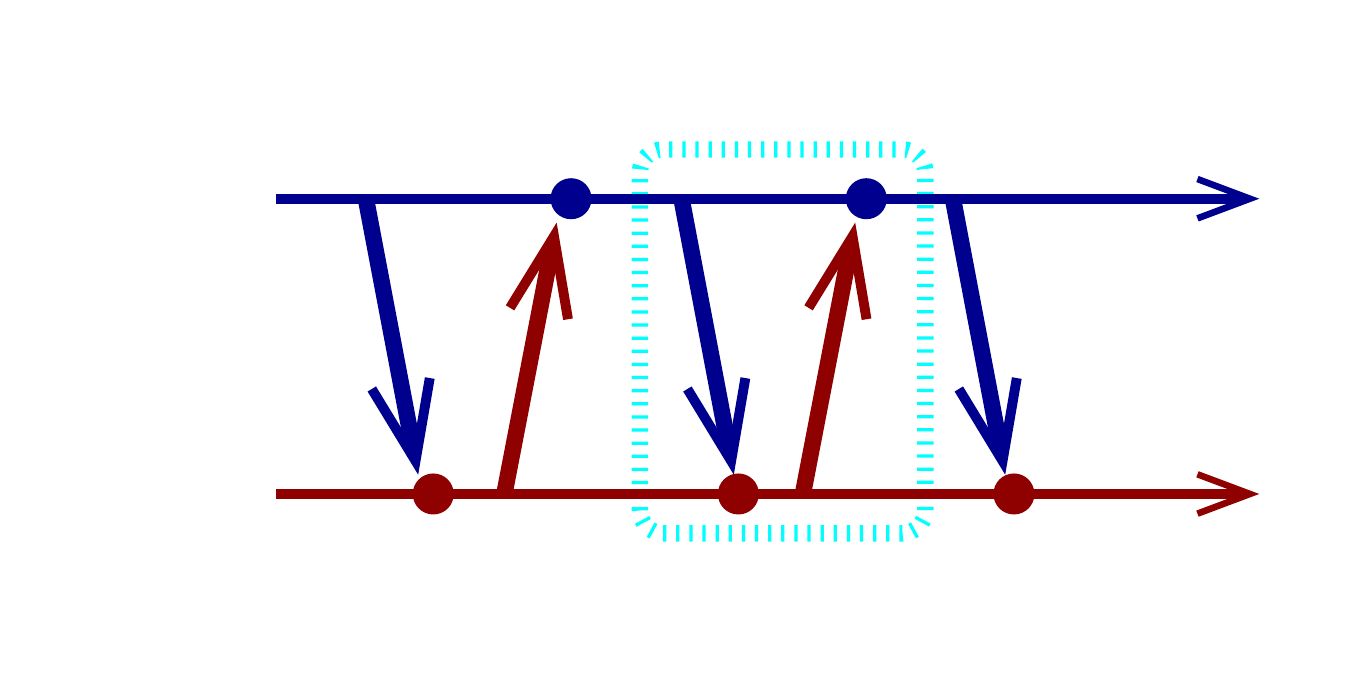}
    \qquad
    \resizebox{0.45\columnwidth}{!}{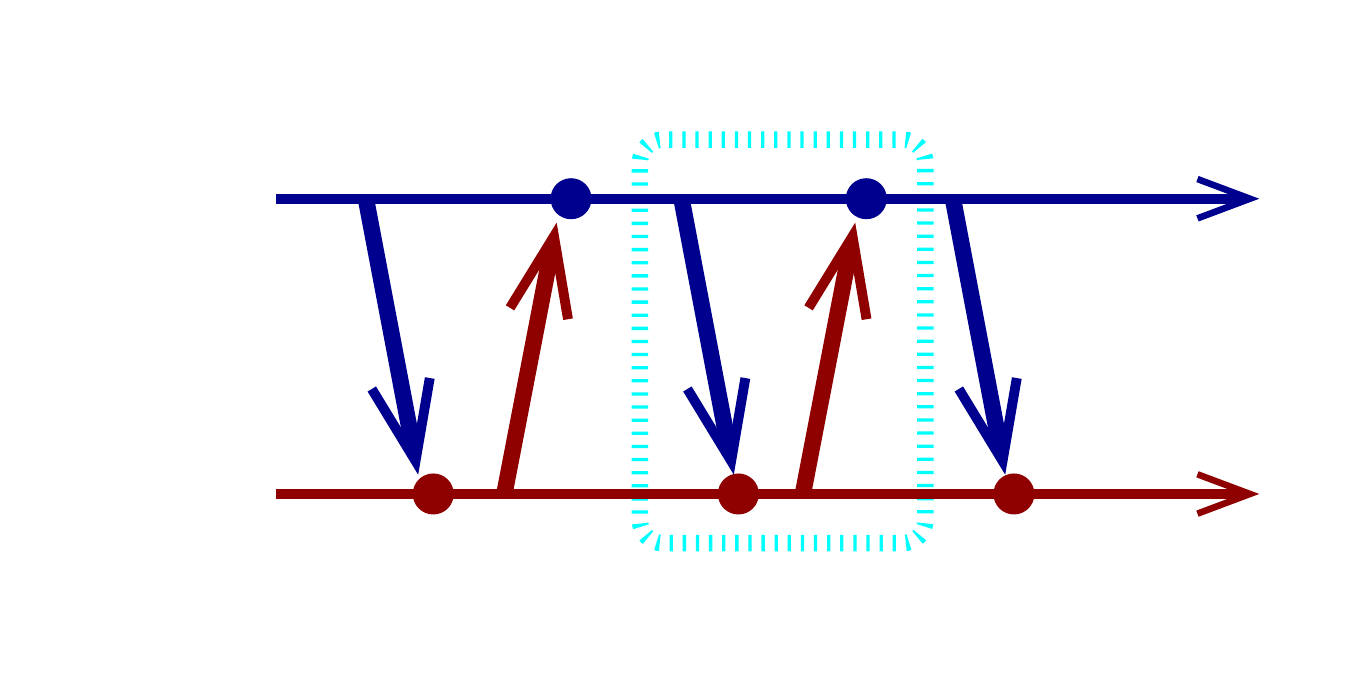}
  \end{center}
\caption[]
{ \label{fig-xvoscil}
A diagram for a simple oscillating system 
 such as a mass on a spring or a pendulum.
The left hand diagram shows the system based on displacement $x$
 and velocity $v$, 
 where the oscillation frequency depends on $F$; 
 the right hand diagram shows the system based on displacement $x$
 and momentum $p$, 
 where the oscillation frequency depends on $M$ and $K$.
}
\end{figure}

However, 
 let us first consider a simple oscillatory system, 
 where the displacement $x$ of some object changes
 according to its velocity $v$, 
 and its velocity $v$ changes according to some (restoring) acceleration
 which is proportional to its displacement.
This is shown on fig. \ref{fig-xvoscil},
 along with an alternate representation 
 based instead on displacement $x$ and momentum $p$, 
 with parameters mass $M$ and restoring force constant $K$.

In the case of causal wave models, 
 we find that spatial variation in the wave amplitude
 drives (causes) changes in the wave profile \cite{Kinsler-2011ejp}.
Although second order wave equations,
 as shown in the previous sections, 
 can be diagrammed rather simply with a single horizontal line, 
 these hide hide half of the dynamics involved in the wave.
From some points of view \cite{Geroch-1996-PDE}
 wave equations exhibiting second order derivatives 
 are merely contracted forms based on a pair of first-order equations, 
 such as in EM or p-Acoustics \cite{Kinsler-M-2014pra}.
Such models have two evolving fields, 
 and hence a pair of horizontal right-travelling lines, 
 where each fields is driven by the spatial derivative 
 of the other.

\begin{figure}[ht]
  \begin{center}
    \resizebox{0.60\columnwidth}{!}{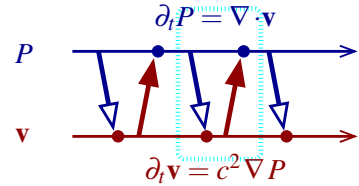}
  \end{center}
\caption[]
{ \label{fig-wave-pA}
A simplified p-Acoustic wave equation, 
 i.e. a sourceless scalar+vector wave equation, 
 consisting of two first-order parts.
We leave the wave-speed squared ``conditioning'' parameter $c^2$
 off the diagram (but not the equations)
 for clarity.
A unit cell is indicated by the dotted oval.
}
\end{figure}

\begin{figure}[ht]
  \begin{center}
    \resizebox{0.60\columnwidth}{!}{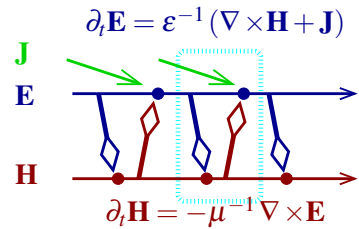}
  \end{center}
\caption[]
{ \label{fig-wave-EM}
The diagram for the vacuum Maxwell's equations, 
 which consist of two first-order equations
 in two the vector fields $\Vec{E}$ and $\Vec{B}$.
We leave the constitutive parameters off the diagram 
 (but not the equations)
 for clarity.
A unit cell is indicated by the dotted oval.
}
\end{figure}

Perhaps the simplest example of this is p-Acoustics \cite{Kinsler-M-2014pra}, 
 a scalar wave theory usually represented by 
 two scalars and two vectors linked by first order differential equations
 and constitutive relations.
To start by keeping things simple, 
 however, 
 we can reduce the model to only two quantities, 
 a scalar population $P$ distribution and a velocity field $\Vec{v}$.
On fig. \ref{fig-wave-pA} we show
 this simplified p-Acoustic model
 with trivial constitutive properties (i.e. only a single speed $c$).
In this model 
 the spatial divergence in the ``flow'' vector field $\Vec{v}$
 cause the ``occupation'' scalar field $P$ to change in time; 
 whilst spatial gradients in the occupation field $P$
 cause the flow field $\Vec{v}$ to change in time.
Likewise on fig. \ref{fig-wave-EM}
 we do the same for a simplified EM, 
 for the electric field $\Vec{E}$ and magnetic field $\Vec{H}$; 
 we see the ordinary curl vector equations
 as influenced by a current source $\Vec{J}$, 
 with only a scalar permittivity $\epsilon$ and permeability $\mu$.

For both figures \ref{fig-wave-pA} and \ref{fig-wave-EM}
 only a small segment of the total time history 
 that would otherwise repeat endlessly into
 the future is shown.
Strictly, 
 only a single subunit or ``unit cell'' --
 as enclosed by the dotted ovals --
 is needed to represent the model.

\begin{figure}[ht]
  \begin{center}
    \resizebox{0.60\columnwidth}{!}{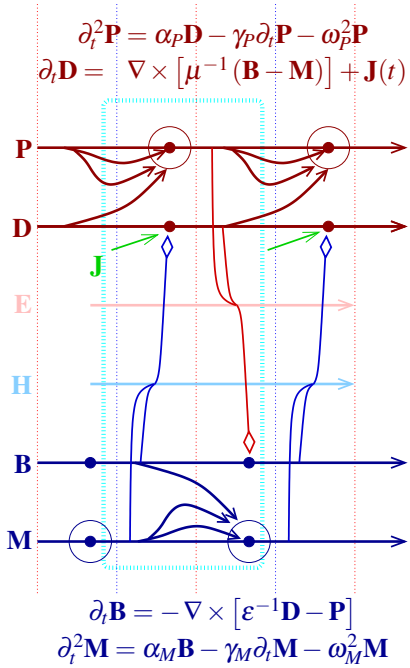}
  \end{center}
\caption[]
{ \label{fig-DPBM-Maxwell}
Causal diagram for the macroscopic Maxwell's equations, 
 with a unit cell being indicated by the dotted oval.
Although fields $\Vec{E}$ and $\Vec{H}$, 
 have been added for clarity, 
 with the arrows from $\Vec{B}, \Vec{M}$ combining to
 indicate how the curl of $\Vec{H}$ causes changes to $\Vec{D}$; 
 and those from $\Vec{D}, \Vec{P}$ combining to
 indicate how the curl of $\Vec{E}$ causes changes to $\Vec{B}$.
No parameters are added to the diagram to avoid clutter, 
 but they are present in the equations.
The locations where the parameters should be added, 
 if so desired, 
 can be inferred from
 the Lorentzian diagram in fig. \ref{fig-selfdecay}, 
 and Maxwell's equations.
}
\end{figure}

A more complete causal diagram for Maxwell's equations
 is shown on fig. \ref{fig-DPBM-Maxwell}.
It mimics the $DPBM$-based FDTD scheme for their numerical solution, 
 and allowing for a Lorentzian response
 for both dielectric polarization $\Vec{P}$
 and magnetization $\Vec{M}$.
Although this representation does not need to include the 
 fields $\Vec{E}$ and $\Vec{H}$, 
 since they follow directly from the in-diagram known 
  $\Vec{D}/\epsilon_0 - \Vec{P}$
 and 
  $\Vec{B}/\mu_0 - \Vec{M}/\mu_0$, 
 it can helpful to add them to the diagram to aid readability, 
 as I have done.

In fact, 
 the link with DPBM numerical scheme
 is not an accident --
 both these diagrams and a computer simulation of a dynamical system
 necessarily must be structured in a causal way.
Computer programs have to calculate things in computational order, 
 and since this is most likely to be in time order, 
 the demands of these diagrams and simulation are the same.
Note, 
 however, 
 that not all dynamical simulations need to chose the time axis 
 to solve along --
 it is common in optics, 
 and sometimes in acoustics,
 to choose a spatial axis along which to integrate
 \cite{Kinsler-2010pra-fchhg,Kinsler-2012arXiv-fbacou}.
In such a case the demands of ``computational causality''
 and traditional physical/temporal causality differ, 
 as has been discussed by Kinsler 
 \cite{Kinsler-2014arXiv-negfreq} (also see references therein).

%
\section{Other Examples, With Complications}
\label{S-others}

\begin{figure}[ht]
  \begin{center}
    \resizebox{0.45\columnwidth}{!}{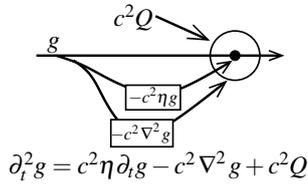}
  \end{center}
\caption[]
{ \label{fig-TDDE}
 Causal diagram for the TDDE.
}
\end{figure}

%
\subsection{The time-dependent diffusion equation}\label{S-TDDE}

The time-dependent diffusion equation (TDDE) \cite{Arfken-MMfPhy}
 is a 
 second order wave equation
 with a loss term added; 
 it appears in a variety of contexts in physics, 
 including acoustic waves in plasmas 
 or the interstitial gas filling a porous,
 statistically isotropic, 
 perfectly rigid solid \cite{MorseIngard-Acoustics}.
It has a three-dimensional, 
 inhomogeneous form for the velocity potential $g \equiv g(\vec{r},t)$ of
 \cite{Buckingham-2008jasa}, 
 and is typically written as 
~
\begin{align}
  \nabla^2
    g
 -
  c^{-2} 
  \partial_t^2
    g
 -
  \eta
  \partial_t
    g
&=
  Q
.
\label{eqn-TDDE-xt}
\end{align}
Here
 { $Q$ is a source term, 
 such as a driving term or some modification to the wave equation.}
Next, 
 $\eta$ is a positive constant that imparts loss,
 and $c$ is the high frequency speed of sound.
The causal diagram for this is shown on fig. \ref{fig-TDDE}, 
 where the usual equation has been rearranged 
 to conform to the explicitly causal form
 with the highest order time derivative on the left hand side.

%
\subsection{An elastic rod wave equation}\label{S-ERod}

Some mathematical models of physical processes, 
 even rather established ones, 
 are less easy to cast as a causal diagram.
To make this point, 
 I will consider an existing model representing 
 acoustic waves traveling along an infinite, 
 isotropic and elastic cylindrical rod of radius $R$.
Following Murnaghan's free energy model, 
 Porubov has derived a wave equation governing propagation
 of the solitary waves along such a rod
 \cite{Porubov-ANLSWS}, 
 and the propagation and dispersive properties of this model 
 in time propagated and space propagated pictures have been compared
 by Kinsler \cite{Kinsler-2012arXiv-fbacou}.
There is no impediment 
 in this model against the rod having ``auxetic'' parameters\cite{Kolat-MTW-2010pssb}, 
 e.g. where the Poisson's ratio was negative \cite{Lakes-1987s}. 
This ``elastic rod equation'' (ERE) 
 describes the displacement $g \equiv g(x,t)$ 
 with a second order wave equation of the form
~
\begin{align}
  c^2
  \partial_x^2
    g
 -
  \partial_t^2
    g
 +
  b_1
  \partial_t^2
  \partial_x^2
    g
 -
  b_2
  \partial_x^4
    g
 +
  \chi
  \partial_x^2
    g^2
&=
  Q
.
\label{eqn-ERod-xt}
\end{align}
 where $g$ is  the longitudinal displacement in the rod\footnote{For 
 the sake of completeness, 
 the other parameters (following \cite{Kolat-MTW-2010pssb}) are:
~
\begin{align}
  c^2 &= \frac{E}{\rho_0}, 
 ~
  \chi = \frac{\beta}{2\rho_0},
 ~
  b_1 = \frac{\nu\left(\nu-1\right) R^2}{2},
 ~
  b_2 = - \frac{\nu E R^2}{2\rho_0},
\nonumber
\\
  \beta 
&= 
  3E 
 +
  l\left(l-2\nu\right)^3
 +
  4m \left(l-2\nu\right)\left(l+\nu\right)
 +
  6n\nu^2
.
\nonumber
\end{align}
Here
 $\beta$ is the nonlinear coefficient, 
 $E$ and $\nu$ are Young's modulus and Poisson's ratio respectively,
 $l$, $m$, $n$ specify Murnaghan's modulus,
 and $\rho_0$ denotes the density.
Poisson's ratio is typically rather small
 (i.e. $|\nu| < 1$), 
 in which case $b_1$ will be negative.
In contrast $b_2$ can cover a wide range of values, 
 especially if auxetic materials are considered, 
 but usually $\nu, E > 0$,
 so that $b_2 <0$.}.
Here, 
 when we  move the highest order time derivatives
 to the left, 
 we get 
~
\begin{align}
  \partial_t^2
  \left(
    1
   +
    b_1
    \partial_x^2
  \right)
    g
&=
  c^2
  \partial_x^2
    g
 -
  b_2
  \partial_x^4
    g
 +
  \chi
  \partial_x^2
    g^2
 -
  Q
.
\label{eqn-ERod-xt2}
\end{align}

Most notably for my purposes here, 
 this model has a problematic feature:
 there are multiple terms in eqn. \eqref{eqn-ERod-xt}
 that have the highest (second) order time derivatives.
These either have to be combined into a single term, 
 to give us a single quantity that suffers the effects of the others, 
 or the equations must be partitioned into two parts
 each of which is causal in appearance.
However, 
 here there is no straightforward way of achieving this here, 
 because of the presence of the $b_1 \partial_x^2$ component; 
 in simpler cases such as 
 (e.g.) the F-model for the spilt ring resonator, 
 it can be done (see sec. IV of \cite{Kinsler-2011ejp}).
Such terms also occur in other systems, 
 such as the Van Wijngarden's equation
 \cite{Eringen-1990ijes,Kinsler-2012arXiv-fbacou} 
 for waves in bubbly liquids:
 i.e., terms combining derivatives which are 
 both second order time and second order in space 
 (here, $b_1 \partial_t^2\partial_x^2 g$).
Other terms, 
 such as the $\partial_t \grad^2 g$ terms 
 appearing in both the Stokes's \cite{Stokes-1845tcps}
 and Van Wijngarden's equation do not cause difficulties
 because they are only first order in the time derivative, 
 which is not the highest order present.

The best strategy here would be to return to the original derivation 
 of the ERE, 
 and alter it:
 perhaps, 
 for example, 
 there might be an equation for how $\partial_x^2 g$ responds to stimuli, 
 so that eqn. \eqref{eqn-ERod-xt2} could be split up
 in some physically motivated way.
Failing that,
 there are two alternate strategies we might use

\emph{First,} 
 we might decide to use a spatial spectrum or ``$k$-space'' picture instead, 
 replacing the space and time displacement $g\equiv g(x,t)$
 with $\tilde{g}\equiv \tilde{g}(k,t)$.
As part of the conversion, 
 each instance of $\partial_x g$ then becomes a factor of $\imath k g$; 
 the resulting $1+b_1 k^2$ factor on the left hand side can then 
 be divided out.
~
\begin{align}
  \partial_t^2
    \tilde{g}(k,t)
&= 
 \frac{
   -
    c^2 k^2
   -
    b_2
    k^4
  }
  {1 + b_1 k^2}
  \tilde{g}(k,t)
   -
 \frac{
    \chi k^2 
    \left[ \tilde{g}(k,t) \ast \tilde{g}(k,t) \right]
  }
  {1 + b_1 k^2}
\nonumber
\\
&\qquad\qquad
 -
  \frac{Q}
  {1 + b_1 k^2}
.
\label{eqn-ERod-xt-k}
\end{align}
This expression can be diagrammed fairly easily, 
 the only drawback being the more complicated calculations
 required to condition how $g$ affects itself.
Also, 
 that while this spatially non-local $k$-space representation
 can remain temporally causal, 
 it does not respect finite signal speeds \cite{Kinsler-2014arXiv-negfreq}.

\emph{Second,}
 we could reject strict temporal causality, 
 and, 
 as is often done in optics and acoustics \cite{Kinsler-2012arXiv-fbacou,Kinsler-2014arXiv-negfreq}, 
 replace it with what might be called a ``spatial causality'' instead.
This is where we propagate our quantities of interest forward
 along some trajectory in space, 
 rather than forward in time.
Thus, 
 instead of following how a spatial profile $g(x)$
 evolves as time $t$ passes, 
 we instead model how the time history $g(t)$
 evolves as we walk along the $x$ axis.
Since eqn. \eqref{eqn-ERod-xt2} has one spatial dimension
 we can ignore the complications of transverse spatial behaviour, 
 and treat $x$ as if it were the time-like dimension.
Writing $\tau = x/c$
 and $\xi = c t$, 
 we then proceed as if $\tau$ really was the time axis.
The relevant highest order derivative is the $b_2 \partial_x^4 g$ one, 
 but it gets converted to $(b_2/ c^{4}) \partial_\tau^4 g$.
With the rest of the conversion complete we find 
 that the scaled time history $g(\xi)$ follows
~
\begin{align}
  \partial_\tau^4
    g
&=
  \frac{c^4}{b_2}
  \partial_\tau^2
    g
 -
  \frac{c^6}{b_2}
  \partial_\xi^2
    g
 +
  \frac{c^4 b_1}{b_2}
  \partial_\xi^2
  \partial_\tau^2
    g
 +
  \frac{c^2\chi}{b_2}
  \partial_\tau^2
    g^2
 -
  \frac{c^4 Q}{b_2}
.
\end{align}
Unlike the spatial spectrum approach given first, 
 each of the cause terms generated here is relatively simple, 
 however there are five of them rather than just three.

In summary, 
 the attempt to diagram this ERE model in a causal way has raised 
 an important issue: 
 not all model equations have rigorously causal interpretations.
This is perhaps not unsurprising, 
 since many simplified descriptions of physical phenomena are far removed
 from the starting point of their derivation, 
 and the effect of simplifying approximations can be subtle and unexpected.


%
\section{Conclusion}
\label{S-conclusions}

One of the primary benefits of this technique is that 
 in making the diagrams, 
 we are forced to clarify what we mean by ``changes''.
It is not uncommon to hear, 
 in conversations about EM, 
 statements along the lines of 
 ``Maxwell's equations tell us that changes in $\Vec{E}$ cause
 changes in $\Vec{B}$''.
Here we are \emph{forced} to distinguish between 
 changes in time (codified here as ``effects'') 
 and changes in space (codified here as potential ``causes'').
Generally, 
 it is best to restrict the use of the word ``change''
 to denote ``effects'',
 i.e. changes in time.

What we can now say, 
 within a rigorously defined prescription \cite{Kinsler-2011ejp}, 
 ``spatial variation in $\Vec{E}$ causes changes in $\Vec{B}$'', 
 or
 ``spatial variation in $\Vec{H}$ causes changes in $\Vec{D}$'', 
 and even draw a diagram to emphasize the point.
Further, 
 the diagram can leads us to an algorithm for computation 
 in which calculation of effects 
 (changes in the state of the system)
 are driven by causes
 (previously calculated states of the system).

It remains to be seen whether or not causal diagrams
 like these find a useful place in building or analysing 
 physical models.
Diagrams 
 such as Feynman \cite{Wiki-FeynmanDiagrams}
 or Wyld diagrams \cite{Wyld-1961ap,Wiki-WyldDiagrams}
 have their fields of application in field theory or fluid mechanics, 
 as a means of bookkeeping and calculation 
 on perturbation expansions.
Others, 
 like Block diagrams \cite{WikiBooks-BlockDiagrams} 
 or DAGS \cite{Elwert-GCM}, 
 and as already mentioned, 
 are not designed
 with dynamical (time dependent) processes in mind.

Here, 
 however, 
 the intent is for a more general type of diagram
 based simply on representing differential equations in a causal way, 
 and as a side effect, 
 also leading the way towards an algorithm for numerical simulation.
As part of this, 
 they also provide a way to reinforce a rigorously grounded 
 notion of causality in even the simplest the physical models.
Even $F=ma$, 
 trivially rearranged to read $\partial_t\Vec{v}=\Vec{F}/m$
 has a causal diagram, 
 one that matches the very first one presented here:
 the left hand panel of fig. \ref{fig-simple}

%
\begin{acknowledgments}
  I acknowledge financial support from the EPSRC 
  grant number EP/K003305/1, 
 and discussions with 
 Thomas 
 Wong.
\end{acknowledgments}

\bibliography{/home/physics/_work/bibtex.bib}

\newpage

%

\end{document}